# Radiative Cooling II: Effects of Density and Metallicity

Ye Wang[1], G. J. Ferland[1], M. L. Lykins[1], R. L. Porter[2], P. A. M. van Hoof[3], R. J. R. Williams[4]

[1] Department of Physics & Astronomy, University of Kentucky
[2] Department of Physics & Astronomy, University of Georgia
[3] Royal Observatory of Belgium, Belgium
[4] Atomic Weapons Establishment plc, UK

Abstract

This work follows Lykins et al. discussion of classic plasma cooling function at low density and solar metallicity. Here we focus on how the cooling function changes over a wide range of density ($n_H < 10^{12} \text{cm}^{-3}$) and metallicity ($Z < 30 Z_\odot$). We find that high densities enhance the ionization of elements such as hydrogen and helium until they reach local thermodynamic equilibrium. By charge transfer, the metallicity changes the ionization of hydrogen when it is partially ionized. We describe the total cooling function as a sum of four parts: those due to H&He, the heavy elements, electron-electron bremsstrahlung and grains. For the first 3 parts, we provide a low-density limit cooling function, a density dependence function, and a metallicity dependence function. These functions are given with numerical tables and analytical fit functions. For grain cooling, we only discuss in ISM case. We then obtain a total cooling function that depends on density, metallicity and temperature. As expected, collisional de-excitation suppresses the heavy elements cooling. Finally, we provide a function giving the electron fraction, which can be used to convert the cooling function into a cooling rate.

Keyword: ISM: general < Interstellar Medium (ISM), Nebulae, plasmas, atomic processes

# 1   Introduction

The plasma cooling function is an important property in many astrophysics problems, including star formation and the interstellar medium. Calculations of the low-density cooling function, such as Sutherland & Dopita (1993) and Lykins et al. (2013), can apply to, for instance, galaxy evolution (Rees & Ostriker 1977, Safranek-Shrader et al. 2010, Schleicher et al. 2010, Skory et al. 2013), where the range of density and metallicity is not extreme. However, the central regions of Active Galactic Nuclei (AGN) have a very wide range of density and metallicity. Broad lined region clouds can have densities extending to as high as $10^{14}$ cm$^{-3}$ while the accretion disk has even higher densities (Osterbrock & Ferland 2006, hereafter AGN3, Peterson 1997). The metallicities in the central regions of quasars can be as high as $30Z_\odot$ (Hamann & Ferland 1999, Wang et al. 2010, 2011). Investigations of the properties of these regions require cooling functions over a very broad range of density and metallicity. However, we know of no work that considers high densities or metallicty, where the cooling can be affected by changes in the ionization of the gas, or in the cooling efficiency of various lines. Only a few studies (such as Bertone et al.2013) discuss non-solar metallicities.

Here, we build upon the Lykins et al. (2013) work to investigate how the collisional ionization equilibrium (CIE) cooling function depends on metallicity and density over the temperature range $10^4 \text{K} < T < 10^{10} \text{K}$. We will discuss cooling below this temperature range in our next paper. We assume that the gas is time steady and that no external radiation field is present. We provide a cooling function that can be used over a large range of metallicity ($Z < 30Z_\odot$) and density ($n_\text{H} < 10^{12} \text{cm}^{-3}$). We find that the cooling for $10^4 \text{K} < T < 2 \times 10^4 \text{K}$ is quite complicated because the ionization of hydrogen changes at high densities (where excited states become important) and high metallicities (where charge transfer affects the hydrogen ionization). The situation is simpler at higher temperatures, where H is mostly ionized. We therefore show two results. First, an analytical fitting function that balances simplicity and accuracy and could easily be incorporated, for instance, into hydrodynamics codes over the range ($2 \times 10^4 \text{K} < T < 10^{10} \text{K}, Z < 30Z_\odot, n_\text{H} < 10^{10} \text{cm}^{-3}$). Second, we also provide numerical tables of results (Appendix C has samples of these Tables, and the full machine readable tables are available online), which can be used when greater precision is needed, for the full range.

We define the cooling rate and cooling function in Section 2.1. We then describe how we separate the total cooling function into different parts, with metallicity and density dependencies, in Section 2.2. Section 3, 4, and 5 discusses these functions and provides their fit results. Section 7 discusses total the cooling function and provides method to convert the cooling function into the cooling rate without knowing the electron density.

# 2 Model

We calculate the cooling rates with the latest development version of the plasma simulation code CLOUDY, which is last described in Ferland et al. (2013). The improvements made to report the cooling given in this paper will be available in the next release of the stable version. We assume CIE with no external light source (no photoionization or Compton cooling). Then we set the temperature, calculate the equilibrium ionization state, and finally calculate the cooling function.

We use solar abundances from Grevesse et al. (2010). For the other metallicities, we leave the H and He abundances unchanged and scale the metals by a factor of $Z$. The primordial case is approximately just the solar case without metals. In this case, we also neglect Li and B, which only have very small abundances. Therefore the primordial case can be treated as the solar case with zero metallicity. We assume that the He/H ratio in these cases is a constant because Steigman (2012) suggests this ratio only changes slightly (less than 10% from $0Z_\odot$ to $Z_\odot$) and helium cooling is not important in high metallicity cases (as shown below). We also calculate cooling for the ISM case in Appendix B. In this case, we assumed Mathis et al. (1977) silicate and graphite grains, and the gas phase abundances that are taken from Cowie & Songaila (1986), Savage & Sembach (1996), Meyer et al. (1998), Snow et al. (2007), and Mullman et al. (1998).

Some cooling/heating rates are zero even through they may have been calculated by CLOUDY. The Compton cooling/heating rate is zero because it depends on the radiation field striking the gas, while we are considering models without external radiation. Our models are static, so cooling due to motions, such as expansion and advection cooling, are also zero. We do not include molecules or molecular cooling because it barely contributes above $10^4$K, the lowest temperature we consider here. Conduction is not included in calculations of cooling functions because it depends on temperature gradients and can only be computed in a macroscopic model of the environment.

## 2.1 Cooling Function

The volume cooling *rate*, sometimes called the cooling power density, of the cooling processes discussed in this paper can be written as $L_C = \sum_X n_e n_X \alpha_X$ [erg cm$^{-3}$ s$^{-1}$], where $n_e$ is the electron density, $n_X$ is the density of particles that produce the cooling, and $\alpha$ is the rate coefficient of cooling process involving electron and *X*. We refer to the species *X* as a coolant. In general *X* could be an ion, atom, or electron.

Most studies focus on low densities, and get a density-independent cooling *function*, $\Lambda = L_C/(n_e n_H)$ [erg cm$^3$ s$^{-1}$], where $n_H$ is number density of Hydrogen. However, we shall see that this is inaccurate, especially for $T < 10^5$K, due to non-linear changes in the gas properties.

In our calculation, we find that the cooling function depends, when considering a large range of densities and metallicities, on both the density, due to cooling suppression or density-deduced ionization shifts (details are discussed in Appendix A), and the metallicity, because of charge transfer and its effects on the free electron density (details are discussed below in Appendix A). We do not make the simple assumption that cooling only depends on temperature and show that this is not appropriate when higher accuracy is desired. We give functions that depend on density and metallicity to obtain a more accurate cooling function. We will use a separation of variables to derive these functions and discuss their basic format in the following section.

## 2.2 Cooling Function Dependencies

We write the total cooling function as the sum of four terms. The H&He term $\Lambda_{\text{H\&He}}$ contains all the cooling produced by the atoms or the ion of H and He, including the electron-ion bremsstrahlung produced by H and He ions. The metal term $\Lambda_{\text{metal}}$ includes all cooling produced by atoms and ions of elements heavier than He, including the electron-ion bremsstrahlung produced by metal ions. The electron term $\Lambda_{\text{ee}}$ is electron-electron bremsstrahlung cooling, which is important at high temperatures. The grain term $\Lambda_{\text{grain}}$ is the cooling due to grains, which is important at high temperatures, if the grains survive. We split the total cooling function into these four terms because changes in the metallicity will change the density of the heavy elements relative to hydrogen and this will have different effects for each of these terms. A python script to obtain these 4 terms from the CLOUDY output is available online (https://github.com/wangye0206/Cloudy_Helper).

The grain term depends on the grain properties, which can be different for different circumstances. We neglect grain cooling in the main part of this paper, while we will briefly discuss it for ISM conditions in Appendix B.

Neglecting grains, the total cooling function can be written as

$$\Lambda = \Lambda_{\text{H\&He}} + \Lambda_{\text{metal}} + \Lambda_{\text{ee}} \quad (\text{Eq 1})$$

Each of these terms can be written as

$$\Lambda_i(T, n_{\text{H}}, Z) = M_i(T, n_{\text{H}}, Z) \times D_i(T, n_H) \times \Lambda_i(T) \quad (\text{Eq 2})$$

where

$$M_i(T, n_{\text{H}}, Z) = \frac{\Lambda_i(T, n_{\text{H}}, Z)}{\Lambda_i(T, n_{\text{H}}, Z=Z_\odot)} \quad (\text{Eq 3})$$

is metallicity dependence function,

$$D_i(T, n_H) = \frac{\Lambda_i(T, n_H, Z=Z_\odot)}{\Lambda_i(T, n_H = 1 cm^{-3}, Z=Z_\odot)} \quad (\text{Eq 4})$$

is density dependence function,

$$\Lambda_i(T) = \Lambda_i(T, n_{\text{H}} = 1 cm^{-3}, Z = Z_\odot) \quad (\text{Eq 5})$$

is the basic low density and solar abundance cooling function, and $i$ indicates the species. These can be H&He for H&He cooling, metal for metal cooling, and ee for electron-electron bremsstrahlung cooling.

The analytical fits to ~~the~~ these functions provided in the following sections are only valid over the reduced range $2\times10^4\text{K} < T < 10^{10}\text{K}$ and density $n_\text{H} < 10^{10}\text{cm}^{-3}$. This is because the ionization of the gas changes in a complicated way outside this range (the "ionization changes", discussed in Appendix A). We present numerical results over the full range in tables. Appendix C contains sample tables while the full tables are available online. Appendix D gives representative values of our fitting functions.

## 3   The H and He cooling

### 3.1   The H&He Cooling Function at Low Density and Solar Abundance

The H&He cooling is shown in Figure 1 and Table 1 (All tables are in Appendix C). There are 2 peaks in the H&He cooling curve. The first, just below $2\times10^4\text{K}$, is due to the ionization of hydrogen. Below this temperature, hydrogen is predominantly atomic. As the temperature increases, electrons collisionally excite, and eventually ionize, H, due to increasing average electron kinetic energy. This increase in the ionization further increases the electron density, which further increases the cooling. However, when nearly all the hydrogen is ionized ($T > 2\times10^4\text{K}$), the amount of atomic cooling decreases because the atomic hydrogen density decreases more than the electron density increases.

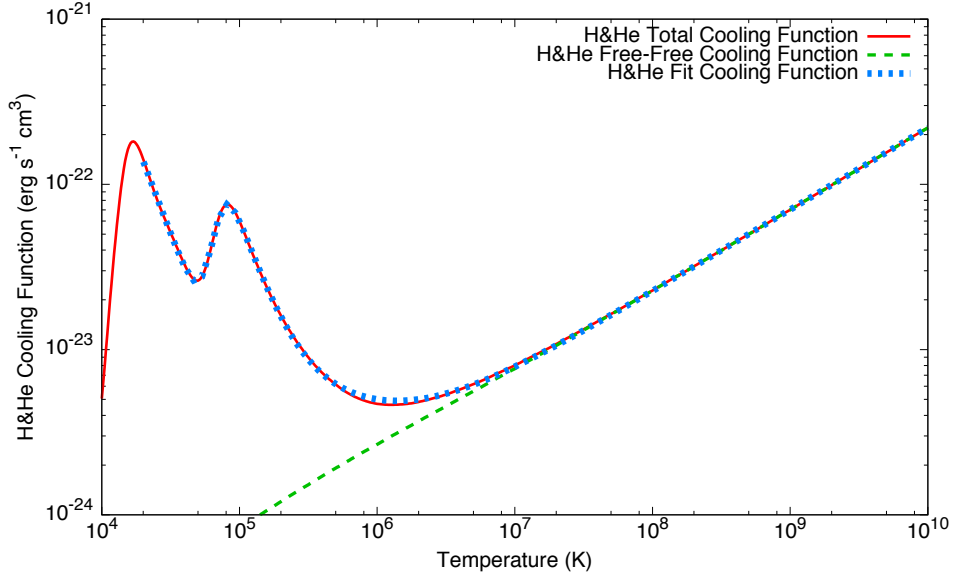

**Figure 1** The H&He Cooling Function $\Lambda_{\text{H\&He}}$. The solid line is the total H&He cooling. The green dashed line is cooling due to H&He free-free emission (bremsstrahlung). The blue dotted line is the fitting function for the total H&He cooling function. The error of this fit is less then 5% above 20,000K. The fit does not apply below this temperature.

Above $5.5 \times 10^4$K the amount of $He^+$ increases and it becomes the leading coolant creating a second peak at $8.5 \times 10^4$K. As the temperature continues to increase, H and He are eventually fully ionized. Both H and He cooling are reduced as a result. Free-free cooling due to electron collisions with $H^+$ and $He^{2+}$ ions is the dominant H&He cooling process for temperatures above $10^6$ K.

The H&He cooling function we fit is shown as the blue line in Figure 1. The function has the form

$$\Lambda_{\text{H\&He}}(T) = \frac{aT^b + (cT)^d(eT^f + gT^h)}{1 + (cT)^d} + iT^j \quad (2 \times 10^4 \text{K} < T < 10^{10}\text{K}) \text{, (Eq 6)}$$

where $a = 4.86567 \times 10^{-13}$, $b = -2.21974$, $c = 1.35332 \times 10^{-5}$, $d = 9.64775$, $e = 1.11401 \times 10^{-9}$, $f = -2.66528$, $g = 6.91908 \times 10^{-21}$, $h = -0.571255$, $i = 2.45596 \times 10^{-27}$, and $j = 0.49521$. The last term in this function is used to describe the behavior of H&He bremsstrahlung cooling. This function reproduces the cooling within 5%.

## 3.2 Density Dependences to H&He Cooling

Figure 2 and Table 2 show the H&He cooling density dependence, which is the cooling relative to the cooling in the low-density limit. At high temperatures ($T \gtrsim 10^6$K), H and He are fully ionized, the bremsstrahlung cooling dominates, and the cooling function does not depend on density, as Figure 2 shows. However, for $T \lesssim 10^5$K, the cooling has a complex dependence on density, with larger variations at higher density. The cooling is suppressed for increasing densities at $1.6\times10^4$K $\lesssim T \lesssim 2\times10^5$K. This is due to the ionization of hydrogen and helium being suppressed as the density increases, as discussed in Appendix A. The result is that the abundances of the leading coolants are reduced and therefore the cooling decreases.

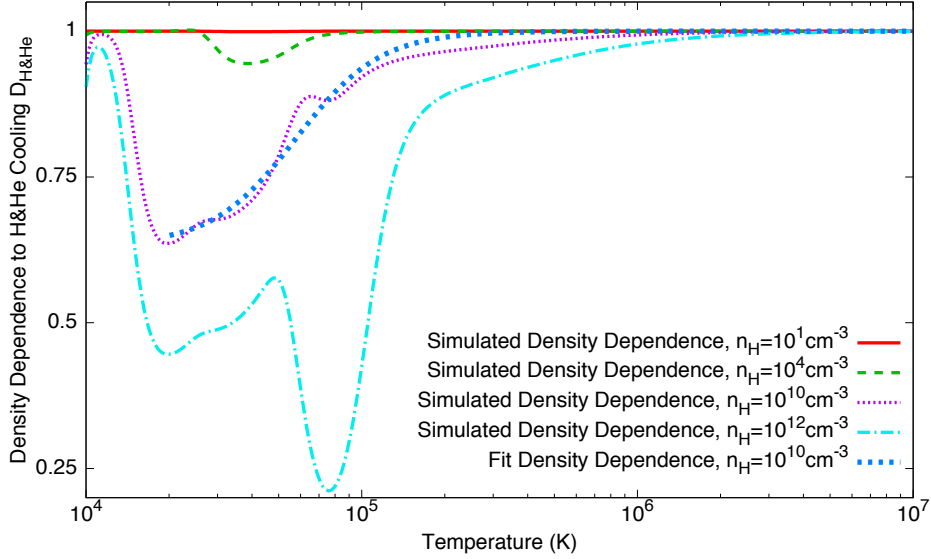

**Figure 2** The H&He Density Dependence Function $D_{\text{H\&He}}$. The large dotted blue line is the fit to the density dependence at $n_\text{H} = 10^{10}$cm$^{-3}$. Other lines show the numerical results at various densities.

The fluctuant features present in Figure 2 are caused by changes in the ionization and cooling coefficient with temperature. We will see similar features in the metal cooling considered below.

We fit the density dependence function for H&He for $T > 2\times10^4$K as

$$D_{\text{H\&He}}(T, n_H) = \frac{T^a + b*g(n_\text{H})}{T^a + b} \qquad (2\times10^4\text{K} < T < 10^{10}\text{K}, n_H < 10^{10}\text{cm}^{-3}), \text{(Eq 7)}$$

where $a = 2.84738, b = 3.62655\times10^{13}, g(n_H) = g_4\times\log^4(n_\text{H}) + g_3\times\log^3(n_\text{H}) + g_2\times\log^2(n_\text{H}) + g_1\times\log(n_\text{H}), g_4 = -3.18564\times10^{-4}, g_3 = 4.8323\times10^{-3}, g_2 = $

$-0.0225974$, and $g_1 = 0.0245446$. The error of this function is less than 5%. We do not fit the wave-like features. The fitting result is shown as the blue line in Figure 2.

### 3.3 Metallicity Dependence to H&He Cooling

We show the H&He cooling function metallicity dependence in Figure 3 and Table 3. Above $10^5$K, the metallicity does not change the H&He cooling, since H&He are generally fully ionized. However, below this temperature, larger metallicity causes increases in the H&He cooling. Charge transfer is responsible for this, as discussed in Appendix A. Increment of the metallicity causes hydrogen and helium to become more neutral. The abundances of the leading coolants ($H^0$ and $He^+$) increase and the cooling is enhanced.

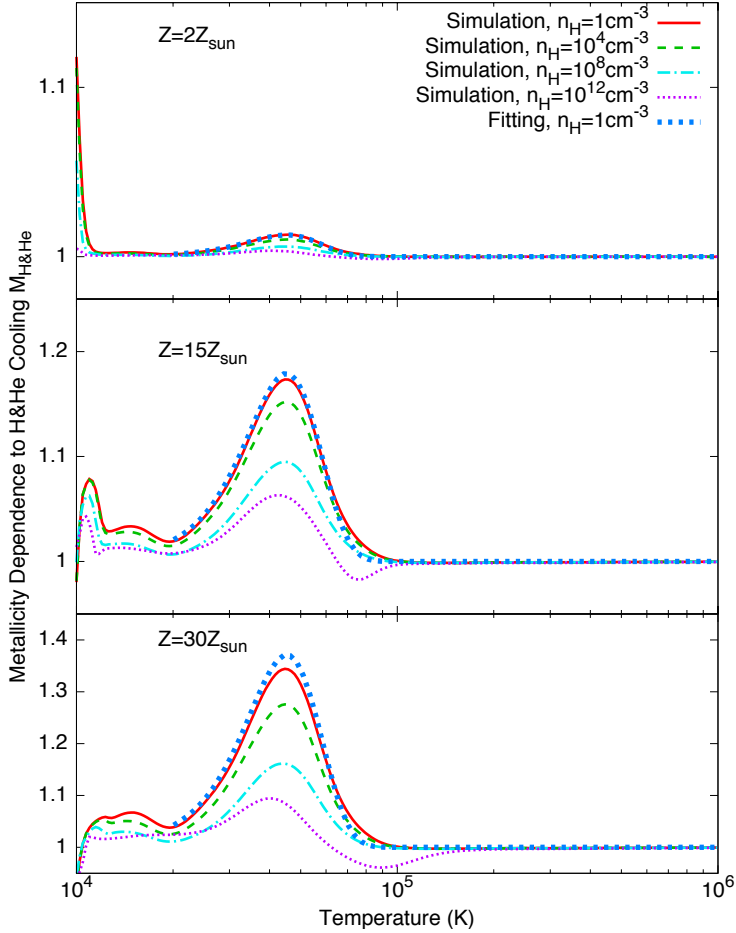

**Figure 3** The H&He Metallicity Dependence $M_{H\&He}$. The three panels show metallicity $2Z_\odot$, $15Z_\odot$, and $30Z_\odot$, from top to bottom. The large dotted blue lines are the fitting function for each metallicity and a density $n_H = 1 cm^{-3}$. The function is equal to one above $10^6$K.

We also find that a density increase partially cancels the effect of metallicity enhancement. This is because increased density suppresses the ionization (Appendix A), partially compensating for the change in the neutral fraction. As a result, metallicity does not enhance the cooling at high density as much as at low density.

We fit the metallicity dependence function for the H&He part as

$$M_{\text{H\&He}}(T, n_H, Z) = (Z-1) \times (a \times \log(n_H) + b) \times \exp\left(-\frac{(T-c)^2}{d}\right) + 1$$

$(2\times 10^4 \text{K} < T < 10^{10}\text{K}, n_H < 10^{10}\text{cm}^{-3}, Z < 30Z_\odot)$ (Eq. 8)

where $a = -0.000847633$, $b = 0.0127998$, $c = 45209.3$, and $d = 2.92145\times 10^8$. The error of this fitting function is less than 5%.

## 4 Metal Cooling

### 4.1 Metal Cooling Function at Low Density and Solar Abundance

The metal cooling as a function of temperature is shown as the red line in Figure 4. This shows the contribution of the metals to the total cooling for solar abundances. As was the case for H&He cooling, bremsstrahlung (the green/dashed line in Figure 4) dominates at high temperatures, where the metals are nearly fully ionized. The numerical result is shown in Table 1.

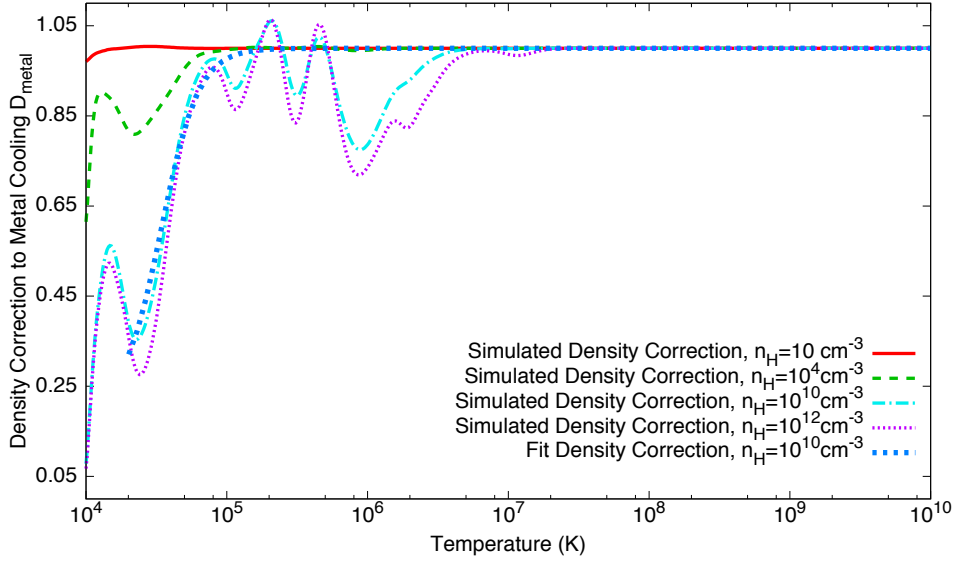

**Figure 4** The Metal Cooling Function $\Lambda_{\text{metal}}$. The solid line is the simulated total metal cooling, the green dashed line presents metal free-free (bremsstrahlung) cooling, while the blue dotted line is our simple fit.

The metal cooling function is complicated because the ionic density and the cooling coefficients change with temperature, which means that the abundance of the leading coolant changes with temperature. This gives the wavy features in the cooling function, which we do not try to fit. Thus, the metal cooling function can be fitted by

$$\Lambda_{\text{metal}}(T) = (aT^b + cT^d)^{-1} + eT^f \quad (2 \times 10^4 \text{K} < T < 10^{10} \text{K}), \text{(Eq 9)}$$

where $a = 6.88502 \times 10^{30}$, $b = -1.90262$, $c = 2.48881 \times 10^{17}$, $d = 0.771176$, $e = 3.00028 \times 10^{-28}$, and $f = 0.472682$. The last term describes metal bremsstrahlung cooling, which is the leading cooling process at high temperatures. This fit is quite approximate, and only reproduces the general shape. At solar metallicity this simplification introduces errors of typically 30% in the total cooling and a maximum 190% error between $4 \times 10^5 \text{K} \sim 5 \times 10^5 \text{K}$.

## 4.2 Density Dependence to Metal Cooling

Figure 5 and Table 4 show the ratio of the cooling at various densities to the cooling at unit density ($1 \text{ cm}^{-3}$), which is low enough for all coolants to be in the low-density limit.

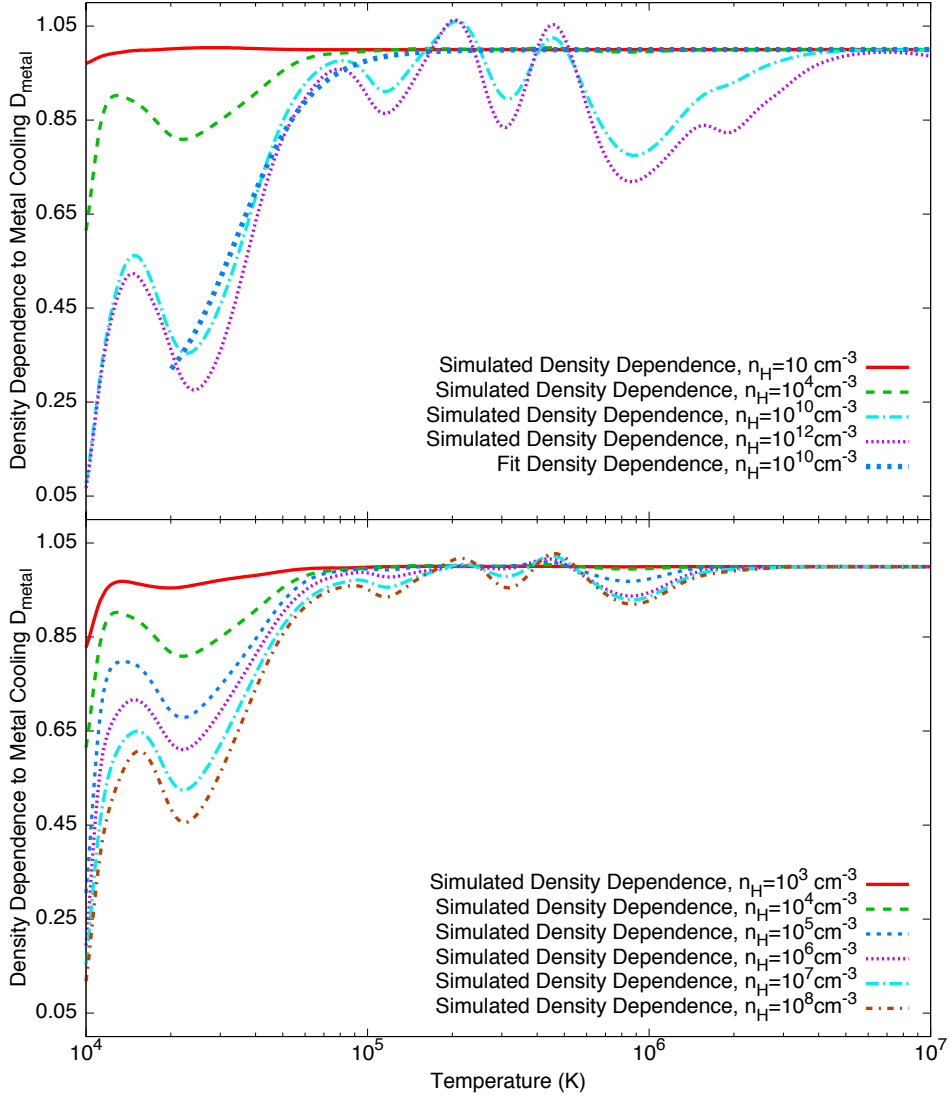

**Figure 5** The Metal Density Dependence $D_{\mathrm{metal}}$. For the upper panel, the large dotted line is the metal-density dependence fit for $n_{\mathrm{H}} = 10^{10}\mathrm{cm}^{-3}$ and the other lines show numerical results. Lower panel shows the density dependence between $n_{\mathrm{H}} = 10^{3}\mathrm{cm}^{-3}$ and $n_{\mathrm{H}} = 10^{8}\mathrm{cm}^{-3}$, where the cooling changes most rapidly.

For most temperatures the cooling is suppressed as the density increases. This is obvious at $T < 10^{7}\mathrm{K}$, and especially at $T < 10^{5}\mathrm{K}$. Much of the cooling over this range is carried by forbidden and inter combination lines.- Forbidden lines are suppressed for densities $n_{\mathrm{e}} \gtrsim 10^{3}-10^{4}\mathrm{cm}^{-3}$ while intercombination lines are suppressed at higher densities, $n_{\mathrm{e}} \gtrsim 10^{8} - 10^{10}\mathrm{cm}^{-3}$.

In addition to the collisional de-excitation suppression effect, the density-induced changes in the ionization discussed in Appendix A also contribute to the changes. This generates the wave-like features in this figure. While collisional de-excitation only suppresses the cooling, the ionization shift can both increase and decrease the cooling, depending on the temperature. For example, the increase in the cooling at high densities around $T\sim 2\times 10^5$K is due to the increase in the ionization of oxygen. The ionization fraction of $O^{4+}$, the leading coolant at this temperature, increases from 42% to 64% as the density increases from $1\text{cm}^{-3}$ to $10^{10}\text{cm}^{-3}$.

We obtain the following density-dependence function for metals

$$D_{\text{metal}}(T, n_\text{H}) = \frac{T^a + b*g(n_\text{H})}{T^a + b} \quad (2\times 10^4\text{K} < T < 10^{10}\text{K}, n_H < 10^{10}\text{cm}^{-3}), \text{(Eq 10)}$$

where $a = 3.29383, b = 8.82636\times 10^{14}, g(n_\text{H}) = g_3\times \log^3(n_\text{H}) + g_2\times \log^2(n_\text{H}) + g_1\times \log(n_\text{H}) + 1$, $g_3 = 0.00221438, g_2 = -0.0353337$, and $g_1 = 0.0524811$. This fits the numerical data to an accuracy of better than 20%.

## 4.3   Metallicity Dependence to Metal Cooling

Changes in the metallicity directly affect the metal cooling. The abundance of the metals scales linearly with Z and therefore increases the cooling. There are additional changes due to interactions between the heavy elements and H and He. We show the function $M_{\text{metal}}(T, n_\text{H}, Z)/Z$ (normalized metallicity dependence for metals) in Figure 6. This function is unity at all parameters except the peak at low temperature ($T \lesssim 1.2\times 10^4$K). This peak is a consequence of charge transfer. As discussed in Appendix A, at these temperatures, high metallicities make hydrogen become more neutral, and hydrogen then neutralizes other elements such as oxygen and carbon. As a result, the abundances of $O^0, Fe^+$, and $C^0$ increase. The total cooling increases as well.

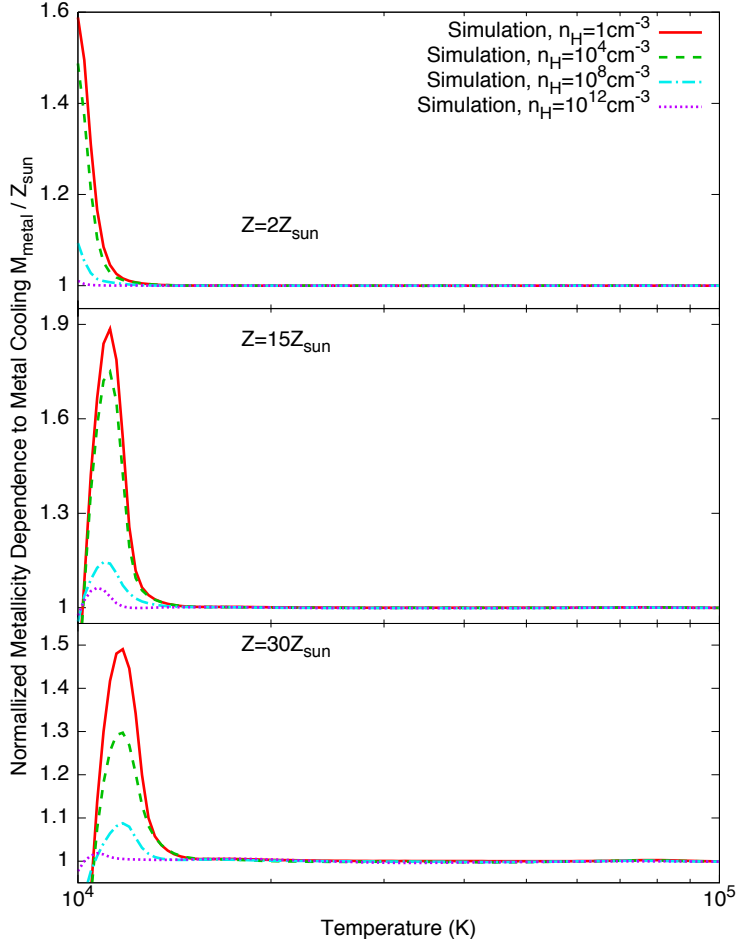

**Figure 6** Normalized Metal Metallicity Dependence $M_{\text{metal}}/Z$. Metallicities of $2Z_\odot$, $15Z_\odot$, and $30Z_\odot$ are shown in the panels from top to bottom. These plots suggest that above $2\times10^4$K the metal metallicity dependence is approximately equal to the metallicity so the ratio is unity.

Table 5 provides results for the full parameter range. Since we only fit the function above $2\times10^4$K, the metallicity dependence for $Z$ can simply be written as

$$M_{\text{metal}}(T, n_H, Z) = Z$$

# 5 The Electron-Electron Bremsstrahlung Cooling

## 5.1 The Electron-Electron Bremsstrahlung Cooling Function at Low Density and Solar Abundances

Electron-electron bremsstrahlung cooling (electron cooling) becomes important at very high temperature ($T \gtrsim 10^9$K), where all atoms are fully ionized and all electrons are in the continuum. We use the electron-electron bremsstrahlung cooling function from Stepney & Guilbert (1983). They calculate the cooling rate for various temperatures in the range $10^8 \text{K} \lesssim T \lesssim 10^{10} \text{K}$. Base on their results, we fit a temperature dependent e-e bremsstrahlung cooling function as

$$\Lambda_{\text{ee}}(T) = \frac{L_{\text{e-e}}}{n_e n_H} = \frac{n_e}{n_H} \sigma_T c \alpha_f \frac{(kT)^2}{m_e c^2} f(T) \quad (2 \times 10^4 \text{K} < T < 10^{10} \text{K}), \text{ (Eq. 11)}$$

where $f(T) = 2.63323 \times 10^3 T^{-0.291936}$ is the fit to the Stepney & Guilbert (1983) data (The lowest temperature in the Stepney & Guilbert data is ~$6 \times 10^8$ K and we extrapolated cooling data to lower temperatures), $\sigma_T$ is the Thomson cross-section, $c$ is the light speed, $\alpha_f$ is the fine structure constant, $k$ is the Boltzmann constant, and $m_e$ is the electron rest mass.

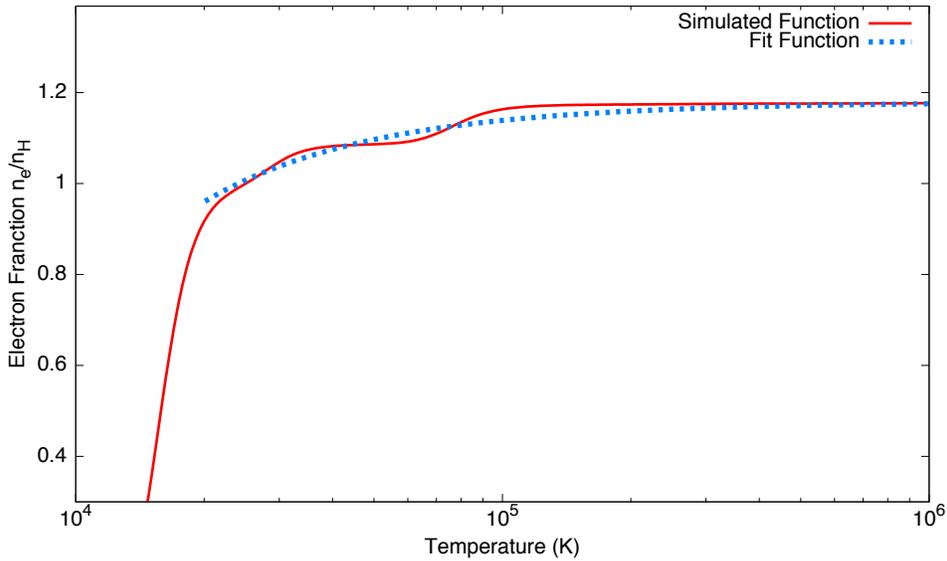

**Figure 7** The electron fraction $n_e/n_H$ in the low-density limit and solar metallicity. The fraction goes to a constant (1.1792) as the temperature becomes large.

The ratio of electron to hydrogen density is a function of temperature under CIE. We use $E(T) = n_e/n_H$ to represent this electron fraction function in the low density limit and for solar metallicity. Figure 7 shows the calculated values and our fit. Calculated values are listed in Table 1. The fit, with an error less than 5% above $2 \times 10^4$K, is

$$E(T) = \frac{n_e}{n_H} = 2.1792 - \exp\left(\frac{3966.27}{T}\right) \quad (2 \times 10^4 \text{K} < T < 10^{10} \text{K}) \text{ (Eq 12)}$$

The rapid increase of $E(T)$ at low temperature is due to the collisional ionization of hydrogen. The second sharp increase is produced by the ionization of He⁺. The subsequent gradual increase is due to the ionization of heavier elements. At higher temperatures, nearly all elements are fully ionized and $E(T)$ tends to a constant (1.1792).

Combining Eq 11 and Eq 12, we find that $\Lambda_{ee}(T)$ is negligible, comparing with the H&He and the metal cooling function, at intermediate to low temperatures ($T \lesssim 10^9$K). This is because electron-electron collisions are homonuclear and have no dipole moment, and hence are mostly elastic at low energies. It is only at kinetic temperatures approaching the rest mass of the electron, $T>10^9$ K, that radiation is effectively produced. Line cooling is not important at high temperatures because the gas is composed of mostly bare nuclei. This makes the high temperature approximation of $E(T)$ valid in the numerical form of $\Lambda_{ee}(T)$. Thus, we eventually provide the fit electron-electron bremsstrahlung cooling function as

$$\Lambda_{ee}(T) = 1.05244 \times 10^{-38} \times T^{1.708064} \qquad (2\times10^4\text{K} < T < 10^{10}\text{K}) \quad \text{(Eq 13)}$$

Figure 8 compares the fit with the Stepney & Guilbert data.

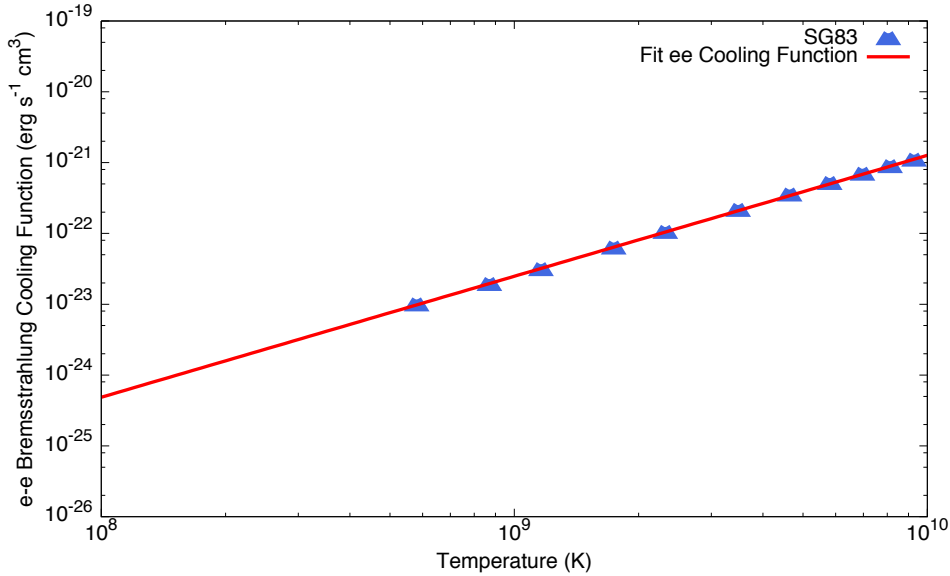

**Figure 8** The e-e Bremsstrahlung Cooling Function $\Lambda_{ee}$. The points give the data from Stepney & Guilbert (1983). The red line is our fit to the SG83 data. We converted the SG83 data, which is originally the cooling rate, to the cooling function by dividing by our derived factor $n_e n_H$ for solar abundances and $n_H = 1\text{cm}^{-3}$.

## 5.2 Density Dependence to Electron Cooling

We show the density dependence $D_{ee}(T, n_H)$ for the electron cooling in Figure 9. Values are also provided in Table 6. This Figure suggests that, below $10^5$K, the ratio is larger than one. This is a result of density-induced changes in the ionization, that is, high densities enhance the ionization, increasing the electron density. Above this temperature the function is near unity because nearly all elements are fully ionized. Therefore the density dependence equals 1.

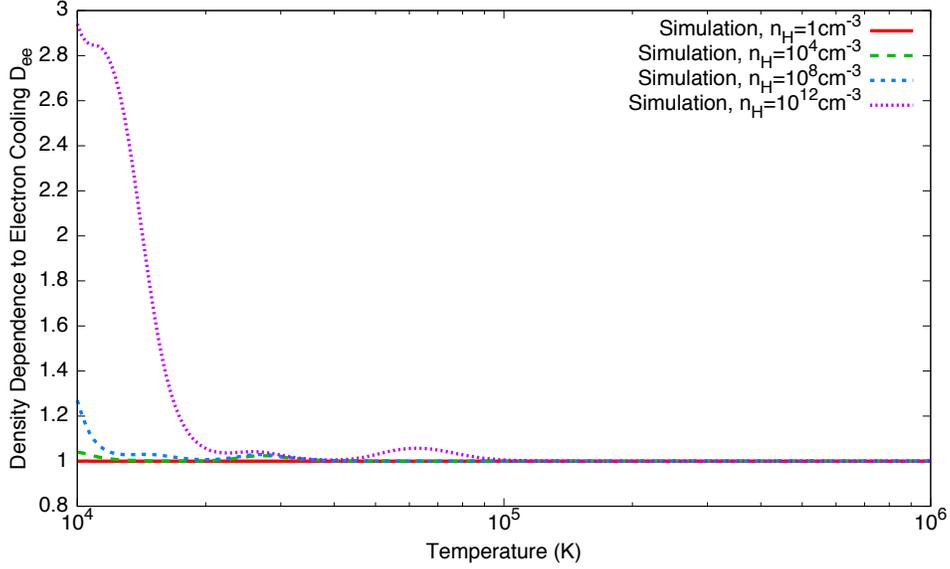

**Figure 9** The e-e Density Dependence $D_{ee}$. For the range we fit, this function is essentially 1 for all relevant temperatures. E-e cooling is only important above $10^8$ K.

We approximate $D_{ee}(T, n_H)$ as unity for $T > 2\times10^4$K, since e-e cooling is only significant when $T > 10^8$ K where this approximation is quite accurate.

### 5.3 Metallicity Dependence to the Electron Cooling

We show the metallicity dependence for electron-electron bremsstrahlung cooling in Figure 10. The ratio has a minimum around $1.2\times10^4$K and the amplitude of this valley decreases with increasing density. This is caused by the density-induced ionization shift described in Section 4 and Appendix A. Above $2\times10^4$K, the metallicity dependence has no density dependence. The ratio increases with the ionization of metals and the amplitude is proportional to metallicity.

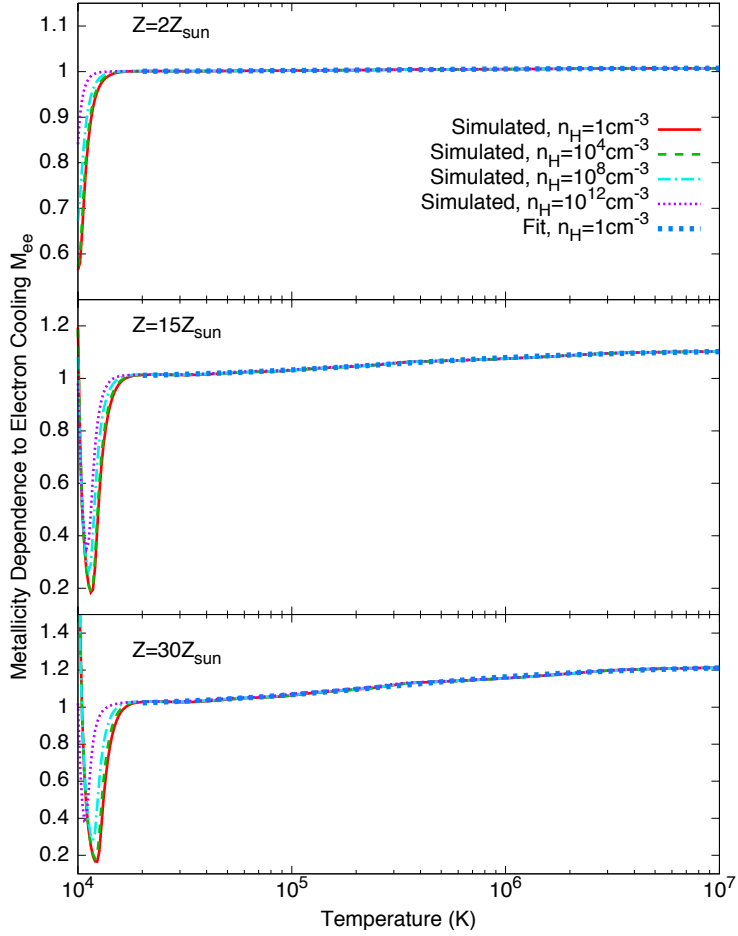

**Figure 10** The H&He Metallicity Dependence $M_{ee}$. The three panels show a metallicity of $2Z_\odot$, $15Z_\odot$, and $30Z_\odot$, from top to bottom. The large dotted blue lines are the fitting functions for each metallicity with density $n_H = 1 \text{cm}^{-3}$.

We fit the metallicity dependence as

$$M_{ee}(T, n_H, Z) = \frac{(a \times Z - a + 1)T^c + b}{T^c + b}$$

$(2 \times 10^4 \text{K} < T < 10^{10} \text{K}, n_H < 10^{10} \text{cm}^{-3}, Z < 30Z_\odot)$ (Eq 14)

where $a = 0.00769985$, $b = 24683.1$, and $c = 0.805234$. The fit result, whose error is less then 1%, is also shown in Figure 10. For high temperatures ($T > 10^7 \text{K}$), where electron-electron bremsstrahlung is important, $M_{ee}(T, n_H, Z) \approx a \times (Z - 1) + 1$. Table 7 gives $M_{ee}$ values with higher precision over a large parameter range.

# 6 The Total Cooling Function and Cooling Rate

## 6.1 The Total Cooling Function

The total cooling function with density and metallicity dependence is

$$\Lambda(T, n_\mathrm{H}, Z) = M_\mathrm{H\&He}(T, n_\mathrm{H}, Z) D_\mathrm{H\&He}(T, n_\mathrm{H}) \Lambda_\mathrm{H\&He}(T)$$
$$+ M_\mathrm{metal}(T, n_\mathrm{H}, Z) D_\mathrm{metal}(T, n_\mathrm{H}) \Lambda_\mathrm{metal}(T)$$
$$+ M_\mathrm{ee}(T, n_\mathrm{H}, Z) D_\mathrm{ee}(T, n_\mathrm{H}) \Lambda_\mathrm{ee}(T) \quad \text{(Eq 15)}$$

Figure 11 shows the calculated cooling function with different densities and metallicities. The metallicity affects the cooling function over the entire temperature range, while changes in the density introduce smaller changes, mainly at $T \lesssim 10^6 \mathrm{K}$. Our density and metallicity dependencies can be applied to cooling functions provided by other work, such as Gnat & Ferland (2012) or Lykins et al. (2013). To do so, simply replace the low density cooling function from this paper with the other cooling functions.

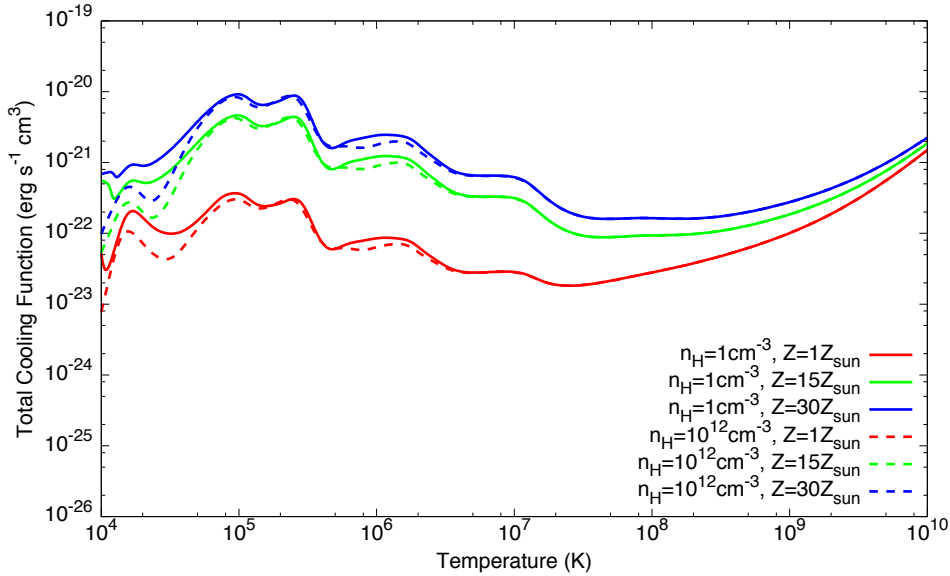

**Figure 11** The cooling function over the range of parameters considered in this paper.

Further simplifications can be done for the analytical fits. From the discussion in Section 4.3, we know that the H&He part of the metallicity dependence function is significantly different from unity between $2 \times 10^4 \mathrm{K}$ to $10^5 \mathrm{K}$ and at high metallicity. The metal cooling dominates the total cooling in these conditions so we can neglect this part of the dependence. The difference between using and neglecting $M_\mathrm{H\&He}(T, n_\mathrm{H}, Z)$ is less than 2%. Combine this and discussions about behaviors of $M_{\text{metal}}(T, n_H, Z)$ and $D_{\text{ee}}(T, n_H)$ in fitting parameter ranges, the final fitting function should be

$$\Lambda(T, n_H, Z) = D_{\text{HHe}}(T, n_H) \times \Lambda_{\text{HHe}}(T)$$

$$+ Z \times D_{\text{metal}}(T, n_H) \times \Lambda_{\text{metal}}(T)$$

$$+ M_{\text{ee}}(T, n_H, Z) \times \Lambda_{\text{ee}}(T) \quad (2 \times 10^4 \text{K} < T < 10^{10} \text{K}, n_H < 10^{10} \text{cm}^{-3}, Z < 30 Z_\odot) \quad (\text{Eq16})$$

## 6.2 From Cooling Function to Cooling Rate

In real astronomy problems, the cooling *rate*, rather than the cooling *function*, is needed. Using the definition of the cooling function, the cooling rate can be obtained as

$$L_C(T, n_H, Z) = n_e n_H \Lambda(T, n_H, Z) = \frac{n_e}{n_H} n_H^2 \Lambda(T, n_H, Z) \quad (\text{Eq 17})$$

To obtain the cooling rate without an independent calculation of the electron density, we need to know $n_e/n_H$ at any density, temperature, and metallicity. Considering the definition of the electron-electron bremsstrahlung cooling function, we can easily write

$$\frac{n_e}{n_H}(T, n_H, Z) = M_{\text{ee}}(T, n_H, Z) D_{\text{ee}}(T, n_H) E(T) \quad (\text{Eq 18})$$

At high temperatures, the electron ratio mainly depends on the metallicity since most elements are fully ionized. At low temperatures, the electron fraction is affected by the density and metallicity, as Figures 9 and 10 show. The electron fraction introduces a strong metallicity and density dependence into the cooling rate.

We combine Eq 17 and Eq 18 to obtain the final cooling function as

$$L(T, n_H, Z) = M_{\text{ee}}(T, n_H, Z) D_{\text{ee}}(T, n_H) E(T) n_H^2 \Lambda(T, n_H, Z)$$

This expression for $n_e/n_H$ can be used with other cooling functions to get the cooling rate.

# 7 Summary

While many calculations of the low density time steady cooling function for solar abundances have been presented in the past, there have been few investigations of how changes in the density or metallicity change the cooling. We summarize our results here.

The largest effect expected from the physics of emission line formation would be the collisional suppression of lines when the density exceeds the critical density of the upper level. This is quite important for regions of the cooling function that are dominated by forbidden transitions, $10^4$ K $< T <10^6$ K. Continuous emission, which is not collisionally suppressed, is the dominant coolant at higher temperatures.

Changes in the density also affect the cooling by changing the ionization of the gas. This is mainly due to processes affecting excited states of H and He, and which bring those species into LTE at high densities. This changes the electron density at temperatures where H or He is partially ionized, which then directly changes the cooling rate. This is especially important for $T < 10^5$ K.

Changes in the metallicity cause direct changes in the cooling due to the changing abundances of the heavy elements. This would be expected to be a simple linear scaling factor. However we find that coupling between the heavy elements and H, He, introduced by charge exchange, causes the H and He ionization to be affected by changes in $Z$. This in turn causes the electron density to change, causing changes in the cooling rate. At high temperatures and $Z$ the metals contribute to the free electron density, further increasing the cooling rate.

Our original goal was to provide simple fits to these dependencies. We find that the physics is quite complex when precision is required. We do present simple analytical predictors of the cooling function and its changes with $T$, $n_\mathrm{H}$, and $Z$, although these are approximate to various degrees. We provide a C and a Python routine that evaluates this cooling function (https://github.com/wangye0206/Cloudy_Helper). Tables giving numerical values of the cooling function as a function of these parameters are also given. Interpolation on these tables should be somewhat more accurate. A method to convert the cooling function into a cooling rate is also provided.

# 8 Acknowledgments

We thank Prof Pete Storey for his careful review and valuable comments. We acknowledge support from NSF (1108928 and 1109061), NASA (10-ATP10-0053, 10-ADAP10-0073, and NNX12H73G), and STScI (HST-AR-12125.01, GO-12560, and HST-GO-12309. PvH acknowledges support from the Belgian Science Policy office through the ESA PRODEX program.

## 10  Appendix A – Density and Metallicity Induced Ionization Shift

In the simplest collisionally ionized case the ionization distribution is only a function of temperature. The ionization balance equations becomes

$$n_e n_{I^0} C = n_e n_{I^+} \alpha \qquad \text{(Eq A1)}$$

where $n_e$ is the electron density, $I$ stands for an ion or atom, $C$ is the collisional ionization rate coefficient from the ground state, and $\alpha$ is the radiative recombination rate coefficient (erg cm$^3$ s$^{-1}$) for all states. We can see that the electron density cancels out. Thus, when the plasma reaches ionization equilibrium, the ionization depends on the ratio of the two rate coefficients, which are functions of temperature.

However, in our calculations, we find that both density and metallicity change the hydrogen ionization, the electron fraction, and the ionization of the heavy elements, by a significant amount. We call this change the ionization shift. This is especially important at the lower end of the temperature range discussed in this paper. Here we discuss the physical reasons why the density and metallicity have an influence on the ionization.

### 10.1 Ionization Shift due to Density

The ionization of hydrogen over a very wide range of density has been considered by Bates et al. (1964) and Summers (1972, 1974). Here we use Cloudy to demonstrate how the ionization changes because of density.

We consider the simplest model, a pure hydrogen plasma at a temperature of $T = 12600$ K and a range of densities. The red solid line in Figure A1 shows our computed ionization of hydrogen as a function of density.

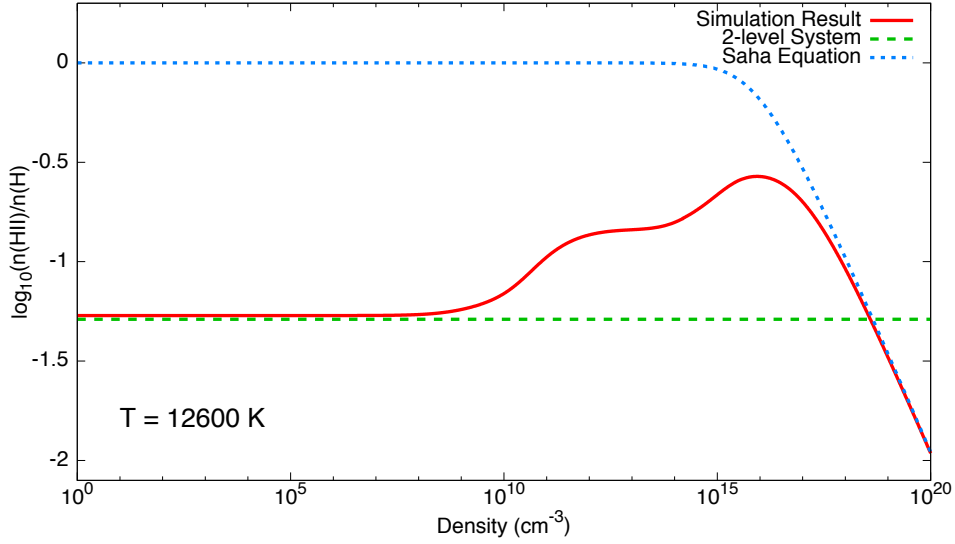

**Figure A** 1 Ionization Shift Due to Density. The solid red line is the numerical result, the green dashed line is the 2-level system result, and the blue dotted line is the result given by the Saha Equation.

The ionization is nearly constant at densities lower than $n_H = 10^8 \text{cm}^{-3}$. Over this range the leading contributors to the ionization and recombination rates are electron collisional ionization from the ground state and radiative recombination. This becomes a "two level system", with only these two terms. Considering the fact that the electron density cancels out, the solution is

$$\frac{n_{H^+}}{n_{H^0}} = \frac{C}{\alpha}. \qquad \text{(Eq A2)}$$

The green line in Figure A1 shows this solution. The ionization and recombination rate coefficients used here come from Voronov (1997) and AGN3, respectively. This is clearly a good approximation at low densities, where Eq A2 overlaps with our complete simulations, although it fails when $n_H > 10^8 \text{ cm}^{-3}$.

The gas is in thermodynamic equilibrium (TE) in the high-density limit. In TE the ionization is described by the Saha equation, which, for pure hydrogen, is

$$\frac{n_{H^+}}{n_{H^0}} = \frac{1}{n_e}\left(\frac{2\pi m_e kT}{h^2}\right)^{\frac{3}{2}} e^{-\frac{13.6\text{eV}}{kT}}. \qquad \text{(Eq A3)}$$

where $m_e$ is electron mass and $h$ is Planck constant. This is shown as the blue dotted line in Fig A1. In this limit the ionization is inversely proportional to the electron density. The calculations show that our computed ionization goes over to this TE limit when $n_H > 10^{15} \text{ cm}^{-3}$. As the figure shows, the ionization is suppressed at high density.

CLOUDY actually solves for the ionization using model hydrogen and helium atoms with many bound levels. The computed ionization is seen to smoothly vary between the asymptotic low and high-density limits. At intermediate densities, $10^8$ cm$^{-3} < n_H < 10^{15}$ cm$^{-3}$, the ionization lies between these two limits. In this range, the increasing density enhances both the ionization and recombination coefficients. The ionization rate increases because of the increasing importance of collisional ionization from excited levels. On the other hand, 3-body recombination also becomes important, and thus increases the total recombination coefficient. At intermediate densities, the extra ionization dominates recombination and the ionization increases as a result.

The ionization of the heavy elements is also found to depend on density, but by a smaller amount. These are treated assuming the effective two-level atom described above. Unlike hydrogen, most heavy elements recombine by dielectronic recombination (AGN3). This process is collisionally suppressed at high densities (Davidson 1975). The decrease in the recombination rate produces a small increase in the ionization of the heavy elements at higher densities, as suggested by eqn A1.

## 10.2 Ionization Shift due to Metallicity

The ionization is also affected by changes in the metallicity. Charge transfer with hydrogen plays an important role in the ionization equilibrium, especially for $T = 1 \times 10^4 \sim 2 \times 10^4$ K. For simplicity, consider the temperatures where gas consists almost entirely of atoms and first ions. The ionization-equilibrium equation of hydrogen can be written as

$$n_{H^0} n_e C + \sum_I n_{H^0} n_{I^+} C_X^i = n_{H^+} n_e \alpha + \sum_I n_{H^+} n_{I^0} C_X^r, \qquad \text{(Eq A4)}$$

where e stands for electrons, $C_X^i$ is hydrogen charge transfer ionization rate coefficient, $C_X^r$ represents hydrogen recombination rate coefficient due to charge transfer and $I^0$ and $I^+$ are the atom and ion of a heavy element, respectively. In this equation, we see, when charge transfer ionization and recombination are comparable with other sources of ionization and recombination, hydrogen's ionization will be sensitive to the heavy-element abundance and ionization.

In our case, tests show that hydrogen's ionization changes due to charge exchange with magnesium by the reaction $Mg^0 + H^+ \rightarrow Mg^+ + H^0$. We use charge transfer rates from Kingdon & Ferland (1996). For the 30 $Z_\odot$ and $T$ = 12,600 K case, the rate of hydrogen charge transfer is $n_{I^0}C_X^r \sim 7\times 10^{-14}$ s$^{-1}$, while the hydrogen radiative recombination rate is $n_e \alpha \sim 1.5\times 10^{-14}$ s$^{-1}$. This means that charge transfer causes hydrogen to become more neutral, as Figure A2 shows. Most of the other elements, such as O, couple with hydrogen, so the plasma ionization changes.

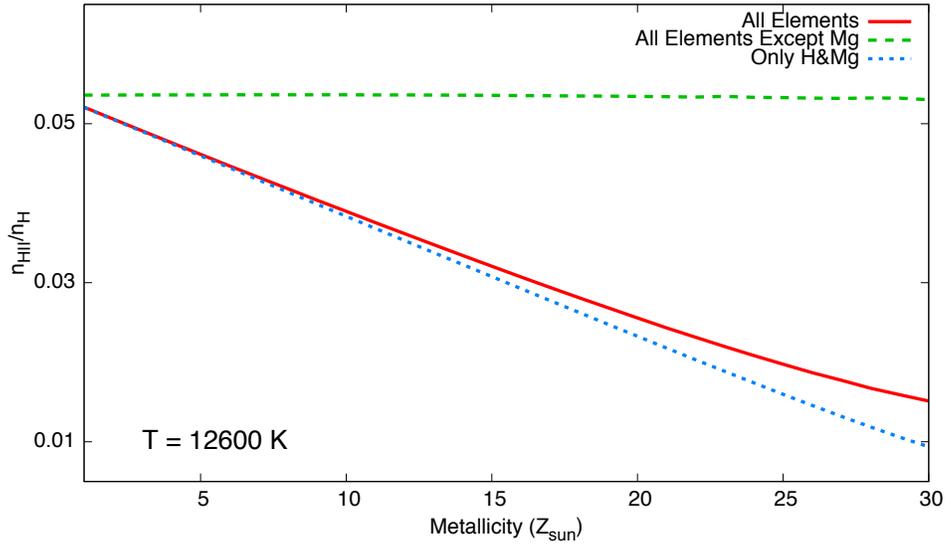

**Figure A 2** The Ionization Shift due to Metallicity.

The importance of Mg is a surprising result because the Mg charge exchange rate is significantly smaller than the resonant O – H charge exchange rate (AGN3). O-H charge exchange goes in both directions, causing ionization and neutralization, so the O ionization couples to that of H, without the H ionization being strongly affected.

# 11 Appendix B Grain Cooling

Grain cooling only occurs in dusty environments. Its amount depends on the grain abundance. We only show grain cooling for the ISM case with depleted gas phase abundances and Mathis et al. (1977) grains. The gas phase abundances are taken from Cowie & Songaila (1986), Savage & Sembach (1996), Meyer et al. (1998), Snow et al.

(2007), and Mullman et al. (1998). The grain cooling rate is discussed in Ostriker & Silk (1973), Draine & Sutin (1987), and van Hoof et al. (2001, 2004).

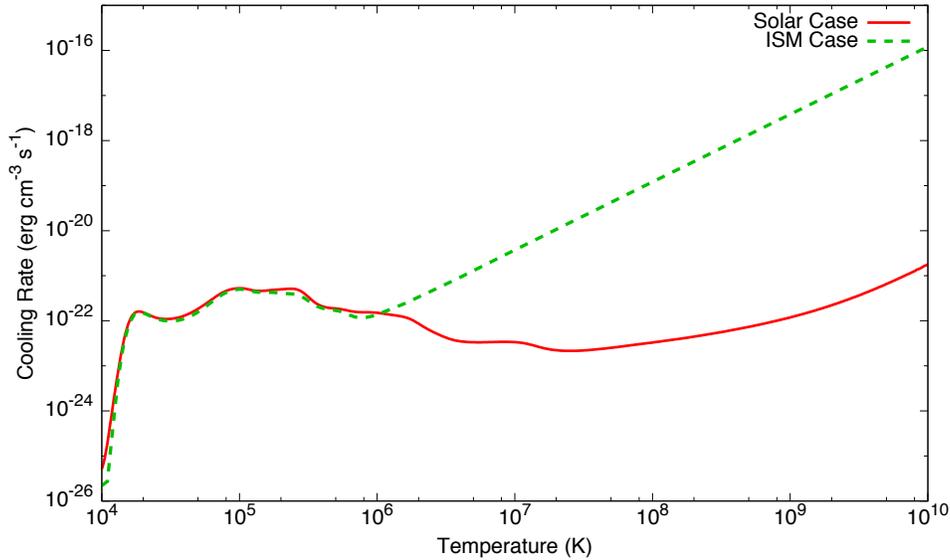

**Figure B 1** Total Cooling Rate for the Solar (red/solid line) and ISM (green/dashed line) Cases.

Figure B1 compares the cooling *rate* for the solar and ISM cases. The solar case has all elements in the gas phase and a solar composition. We also assume a hydrogen density of $1~\text{cm}^{-3}$. Because of depletion of some heavy elements in ISM case, the gas cooling is decreased between $2\times10^4$K and $10^6$K. These are the temperatures where metal line radiation dominates the total cooling. Above $10^6$K, the grains become the dominant cooling.

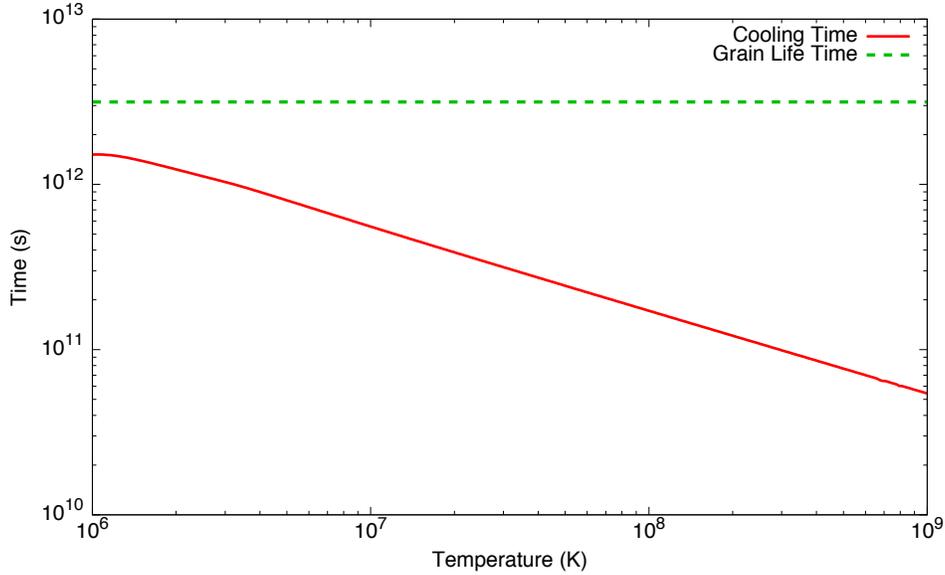

**Figure B 2** Cooling time and grain life time. Grains are assumed to have a radius of 0.05 μm and hydrogen density is 1 cm⁻³.

We do not provide a fitting function for ISM abundances; table 8 is used to provide the cooling *rate* of the ISM case. The grains actually may not be able to survive a long time when temperature is high. Draine & Salpeter (1979) suggest that the lifetime of grains, for $T > 10^6$K, is of the order $\sim 2\times10^4 \times (1/n_\mathrm{H}) \times (a/0.01)$ yr, where $n_\mathrm{H}$ is hydrogen density in cm⁻³ and $a$ is grain radius in μm. For lower temperatures, this timescale is much longer. We assumed a grain radius of 0.05 μm. Figure B2 compares the cooling time and grain lifetime at $n_\mathrm{H} = 1 \mathrm{cm}^{-3}$ for $T > 10^6$K, where the grain cooling becomes important. The grain lifetime is 2~50 times longer than the cooling timescale. Our ISM cooling rate can be applied to problems where the time scales are shorter than the grain lifetime. For timescales that are longer than the grain lifetime, the grain-free cooling function should be used.

# 12 Appendix C Tables

Table 1 lists the electron fraction, and the three different contributors (H&He, metal, and electron-electron bremsstrahlung) to the cooling function, at $n_{\rm H} = 1$ cm$^{-3}$ and $Z_\odot$, in column 2, 3, 4 and 5, respectively.

Table 1.  Cooling Function and Electron Fraction at $n_{\rm H}=1$ cm$^{-3}$ and $Z_\odot$

| Temperature (K) | Electron Fraction $E(T)$ | H&He Cooling $\Lambda_{\rm H\&He}(T)$ (erg cm$^3$ s$^{-1}$) | Metal Cooling $\Lambda_{\rm metal}(T)$ (erg cm$^3$ s$^{-1}$) | Electron Cooling $\Lambda_{\rm ee}(T)$ (erg cm$^3$ s$^{-1}$) |
|---|---|---|---|---|
| 10000 | 9.7504×10$^{-04}$ | 5.0548×10$^{-24}$ | 4.7286×10$^{-23}$ | 5.9103×10$^{-35}$ |
| 10233 | 1.6158×10$^{-03}$ | 6.2325×10$^{-24}$ | 3.3227×10$^{-23}$ | 1.0187×10$^{-34}$ |
| 10471 | 2.6029×10$^{-03}$ | 7.8148×10$^{-24}$ | 2.5198×10$^{-23}$ | 1.7069×10$^{-34}$ |
| 10715 | 4.0692×10$^{-03}$ | 9.8813×10$^{-24}$ | 2.0659×10$^{-23}$ | 2.7755×10$^{-34}$ |
| 10965 | 6.1863×10$^{-03}$ | 1.2513×10$^{-23}$ | 1.8126×10$^{-23}$ | 4.3887×10$^{-34}$ |
| 11220 | 9.1914×10$^{-03}$ | 1.5823×10$^{-23}$ | 1.6744×10$^{-23}$ | 6.7822×10$^{-34}$ |
| 11482 | 1.3396×10$^{-02}$ | 1.9938×10$^{-23}$ | 1.6052×10$^{-23}$ | 1.0281×10$^{-33}$ |
| 11749 | 1.9201×10$^{-02}$ | 2.4967×10$^{-23}$ | 1.5797×10$^{-23}$ | 1.5328×10$^{-33}$ |
| 12023 | 2.7112×10$^{-02}$ | 3.1040×10$^{-23}$ | 1.5832×10$^{-23}$ | 2.2511×10$^{-33}$ |
| 12303 | 3.7747×10$^{-02}$ | 3.8318×10$^{-23}$ | 1.6069×10$^{-23}$ | 3.2599×10$^{-33}$ |

Note. — Full table is provided online as machine readable table.

Table 2 is the density-dependent function to H&He cooling from $n_{\rm H} = 10$ cm$^{-3}$ to $n_{\rm H} = 10^{12}$ cm$^{-3}$.

Table 2. H&He Cooling Density Dependence $D_{\text{H\&He}}(T, n_{\text{H}})$

| Temperature (K) | n=1 | n=2 | ... | n=12 |
|---|---|---|---|---|
| 10000 | 1.0000 | 0.9998 | ... | 0.9038 |
| 10233 | 1.0000 | 0.9999 | ... | 0.9371 |
| 10471 | 1.0002 | 0.9999 | ... | 0.9571 |
| 10715 | 1.0000 | 0.9999 | ... | 0.9675 |
| 10965 | 1.0000 | 0.9999 | ... | 0.9722 |
| 11220 | 1.0000 | 0.9999 | ... | 0.9723 |
| 11482 | 1.0000 | 0.9999 | ... | 0.9684 |
| 11749 | 1.0000 | 0.9999 | ... | 0.9605 |
| 12023 | 1.0000 | 0.9999 | ... | 0.9486 |
| 12303 | 1.0000 | 1.0000 | ... | 0.9322 |

Note. — n = $\log(n_{\text{H}})$, $n_{\text{H}}$ is in unit of cm$^{-3}$. Full table is provided online as machine readable table.

Table 3 is the metallicity dependence to H&He cooling from $2Z_\odot$ to $30Z_\odot$. The data can be extrapolated to $0Z_\odot$. In this table, column 2 is density, and columns 3 to 10 are data for different metallicities.

Table 3.  H&He Cooling Metallicity Dependence $M_{\rm H\&He}(T, n_{\rm H}, Z)$

| Temperature (K) | $n$ | $Z=2Z_\odot$ | $Z=6Z_\odot$ | $Z=10Z_\odot$ | $Z=14Z_\odot$ | $Z=18Z_\odot$ | $Z=22Z_\odot$ | $Z=26Z_\odot$ | $Z=30Z_\odot$ |
|---|---|---|---|---|---|---|---|---|---|
| 10000 | 0 | 1.1180 | 1.1286 | 1.0344 | 0.9884 | 0.9630 | 0.9445 | 0.9315 | 0.9218 |
| 10233 | 0 | 1.0699 | 1.1768 | 1.0856 | 1.0406 | 1.0137 | 0.9959 | 0.9833 | 0.9738 |
| 10471 | 0 | 1.0334 | 1.1449 | 1.1148 | 1.0719 | 1.0455 | 1.0283 | 1.0160 | 1.0067 |
| 10715 | 0 | 1.0149 | 1.1184 | 1.1063 | 1.0795 | 1.0638 | 1.0473 | 1.0353 | 1.0265 |
| 10965 | 0 | 1.0076 | 1.0764 | 1.0980 | 1.0828 | 1.0671 | 1.0546 | 1.0450 | 1.0374 |
| 11220 | 0 | 1.0044 | 1.0382 | 1.0785 | 1.0786 | 1.0691 | 1.0597 | 1.0517 | 1.0452 |
| 11482 | 0 | 1.0030 | 1.0199 | 1.0510 | 1.0676 | 1.0667 | 1.0614 | 1.0557 | 1.0505 |
| 11749 | 0 | 1.0024 | 1.0133 | 1.0295 | 1.0500 | 1.0596 | 1.0602 | 1.0578 | 1.0545 |
| 12023 | 0 | 1.0021 | 1.0111 | 1.0211 | 1.0338 | 1.0472 | 1.0554 | 1.0578 | 1.0574 |
| 12303 | 0 | 1.0021 | 1.0105 | 1.0189 | 1.0278 | 1.0376 | 1.0473 | 1.0545 | 1.0587 |

Note. — n = $\log(n_{\rm H})$, $n_{\rm H}$ is in unit of cm$^{-3}$. Full table is provided online as machine readable table and it contains data for all densities and metallicities considered in this paper.

Table 4 is the density dependence function to the metal cooling. The format of this table is as same as Table 2.

Table 4. Metal Cooling Density Dependence $D_{\mathrm{metal}}(T, n_{\mathrm{H}})$

| Temperature (K) | n=1 | n=2 | ⋯ | n=12 |
|---|---|---|---|---|
| 10000 | 0.9706 | 0.9018 | ⋯ | 0.0675 |
| 10233 | 0.9732 | 0.9118 | ⋯ | 0.1040 |
| 10471 | 0.9759 | 0.9233 | ⋯ | 0.1495 |
| 10715 | 0.9792 | 0.9354 | ⋯ | 0.1992 |
| 10965 | 0.9821 | 0.9468 | ⋯ | 0.2482 |
| 11220 | 0.9847 | 0.9566 | ⋯ | 0.2935 |
| 11482 | 0.9868 | 0.9646 | ⋯ | 0.3337 |
| 11749 | 0.9885 | 0.9708 | ⋯ | 0.3690 |
| 12023 | 0.9899 | 0.9757 | ⋯ | 0.3996 |
| 12303 | 0.9911 | 0.9795 | ⋯ | 0.4262 |

Note. — n= $\log(n_{\mathrm{H}})$, $n_{\mathrm{H}}$ in unit of $\mathrm{cm}^{-3}$. Full table is provided online as machine readable table.

Table 5 is the metallicity dependence function to the metal cooling. This table has the same format as table 3.

Table 5. Metal Cooling Metallicity Dependence $M_{\rm metal}(T, n_{\rm H}, Z)$

| Temperature (K) | n | $Z=2Z_\odot$ | $Z=6Z_\odot$ | $Z=10Z_\odot$ | $Z=14Z_\odot$ | $Z=18Z_\odot$ | $Z=22Z_\odot$ | $Z=26Z_\odot$ | $Z=30Z_\odot$ |
|---|---|---|---|---|---|---|---|---|---|
| 10000 | 0 | 3.1769 | 9.0008 | 10.2821 | 11.2604 | 12.2534 | 13.0434 | 13.8040 | 14.5515 |
| 10233 | 0 | 2.9897 | 12.4671 | 14.4008 | 15.8770 | 17.1855 | 18.4063 | 19.5793 | 20.7212 |
| 10471 | 0 | 2.6199 | 13.3573 | 18.7624 | 20.8755 | 22.6487 | 24.3582 | 25.9983 | 27.5948 |
| 10715 | 0 | 2.3321 | 13.5181 | 20.4976 | 24.2371 | 27.7246 | 29.9598 | 32.0518 | 34.1205 |
| 10965 | 0 | 2.1701 | 11.4578 | 21.2370 | 26.5559 | 30.3151 | 33.4587 | 36.3079 | 38.9922 |
| 11220 | 0 | 2.0908 | 8.7022 | 19.5607 | 26.8986 | 31.8487 | 35.8090 | 39.2922 | 42.5181 |
| 11482 | 0 | 2.0521 | 7.1743 | 15.8032 | 24.9823 | 31.4526 | 36.4099 | 40.6156 | 44.4155 |
| 11749 | 0 | 2.0322 | 6.5723 | 12.5277 | 20.9508 | 28.9321 | 35.1421 | 40.2477 | 44.7165 |
| 12023 | 0 | 2.0212 | 6.3197 | 11.1593 | 17.1205 | 24.5195 | 31.8537 | 38.0614 | 43.3774 |
| 12303 | 0 | 2.0145 | 6.1984 | 10.6314 | 15.4671 | 20.9979 | 27.3851 | 34.0227 | 40.2471 |

Note. — n = $\log(n_{\rm H})$, $n_{\rm H}$ is in unit of cm$^{-3}$. Full table is provided online as machine readable table and it contains data for all densities and metallicities considered in this paper.

Table 6 is the density dependence function to the electron-electron bremsstrahlung cooling (electron cooling). The format of this table is the same as Table 2.

Table 6. Electron Cooling Density Dependence $D_{\mathrm{ee}}(T, n_{\mathrm{H}})$

| Temperature (K) | n=1 | n=2 | ⋯ | n=12 |
|---|---|---|---|---|
| 10000 | 1.0000 | 1.0009 | ⋯ | 2.9404 |
| 10233 | 1.0001 | 1.0013 | ⋯ | 2.8923 |
| 10471 | 1.0004 | 1.0018 | ⋯ | 2.8623 |
| 10715 | 1.0001 | 1.0022 | ⋯ | 2.8491 |
| 10965 | 1.0002 | 1.0023 | ⋯ | 2.8459 |
| 11220 | 1.0002 | 1.0024 | ⋯ | 2.8409 |
| 11482 | 1.0002 | 1.0023 | ⋯ | 2.8269 |
| 11749 | 1.0002 | 1.0022 | ⋯ | 2.8003 |
| 12023 | 1.0003 | 1.0020 | ⋯ | 2.7586 |
| 12303 | 1.0003 | 1.0018 | ⋯ | 2.7007 |

Note. — n= $\log(n_{\mathrm{H}})$, $n_{\mathrm{H}}$ in unit of cm$^{-3}$. Full table is provided online as machine readable table.

Table 7 is the metallicity dependence function to the electron-electron bremsstrahlung cooling (electron cooling) cooling. This table has the same format as Table 3.

Table 7. Electron Cooling Metallicity Dependence $M_{\text{ee}}(T, n_{\text{H}}, Z)$

| Temperature (K) | n | $Z=2Z_\odot$ | $Z=6Z_\odot$ | $Z=10Z_\odot$ | $Z=14Z_\odot$ | $Z=18Z_\odot$ | $Z=22Z_\odot$ | $Z=26Z_\odot$ | $Z=30Z_\odot$ |
|---|---|---|---|---|---|---|---|---|---|
| 10000 | 0 | 0.5648 | 0.5458 | 0.8334 | 1.1219 | 1.3881 | 1.6760 | 1.9639 | 2.2501 |
| 10233 | 0 | 0.5811 | 0.3553 | 0.5377 | 0.7205 | 0.9029 | 1.0855 | 1.2679 | 1.4501 |
| 10471 | 0 | 0.6504 | 0.3029 | 0.3563 | 0.4741 | 0.5951 | 0.7141 | 0.8332 | 0.9524 |
| 10715 | 0 | 0.7320 | 0.2596 | 0.2822 | 0.3477 | 0.4029 | 0.4822 | 0.5632 | 0.6432 |
| 10965 | 0 | 0.8018 | 0.2731 | 0.2254 | 0.2588 | 0.3050 | 0.3553 | 0.4073 | 0.4602 |
| 11220 | 0 | 0.8541 | 0.3599 | 0.2062 | 0.2068 | 0.2320 | 0.2639 | 0.2984 | 0.3342 |
| 11482 | 0 | 0.8923 | 0.4903 | 0.2397 | 0.1859 | 0.1894 | 0.2062 | 0.2277 | 0.2513 |
| 11749 | 0 | 0.9202 | 0.6126 | 0.3505 | 0.2094 | 0.1751 | 0.1745 | 0.1841 | 0.1979 |
| 12023 | 0 | 0.9406 | 0.7090 | 0.4924 | 0.3127 | 0.2093 | 0.1726 | 0.1650 | 0.1680 |
| 12303 | 0 | 0.9555 | 0.7816 | 0.6147 | 0.4584 | 0.3233 | 0.2302 | 0.1843 | 0.1655 |

Note. — n = $\log(n_{\text{H}})$, $n_{\text{H}}$ is in unit of cm$^{-3}$. Full table is provided online as machine readable table and it contains data for all densities and metallicities considered in this paper.

Table 8 is the electron fraction and cooling rate at $n_{\text{H}} = 1$ cm$^{-3}$ for ISM abundances with grains in column 2 and 3, respectively.

Table 8. ISM Cooling Rate at $n_{\rm H}$ =1 cm$^{-3}$

| Temperature (K) | Electron Fraction $E_{\rm ISM}(T)$ | ISM Cooling $L_{\rm ISM}(T)$ (erg cm$^{-3}$ s$^{-1}$) |
|---|---|---|
| 10000 | $3.0220\times10^{-05}$ | $2.2146\times10^{-26}$ |
| 10233 | $4.1431\times10^{-05}$ | $2.2914\times10^{-26}$ |
| 10471 | $5.8642\times10^{-05}$ | $2.3869\times10^{-26}$ |
| 10715 | $8.9845\times10^{-05}$ | $2.5192\times10^{-26}$ |
| 10965 | $1.4514\times10^{-04}$ | $2.7224\times10^{-26}$ |
| 11220 | $2.6981\times10^{-04}$ | $3.1255\times10^{-26}$ |
| 11482 | $7.0029\times10^{-04}$ | $4.4430\times10^{-26}$ |
| 11749 | $3.3370\times10^{-03}$ | $1.3227\times10^{-25}$ |
| 12023 | $1.0298\times10^{-02}$ | $4.1916\times10^{-25}$ |
| 12303 | $2.0537\times10^{-02}$ | $9.6560\times10^{-25}$ |

Note. — Full table is provided online as machine readable table.

# 13 Appendix D Fitting Function Check Table

Table D1 can be used to verify the correct evaluation of the fitting coefficients given above. This table gives values of fitting functions at $n_{\rm H} = 10^5 {\rm cm}^{-3}$ and $Z = 15 Z_\odot$.

Table D1. Fitting Functions Check Table

| Temperature(K) | $2 \times 10^4$ | $10^{4.5}$ | $10^5$ | $10^6$ | $10^8$ | $10^{10}$ |
|---|---|---|---|---|---|---|
| $E(T)$ | 0.95986 | 1.0456 | 1.1387 | 1.1752 | 1.1792 | 1.1792 |
| $\Lambda_{\rm H\&He}(T)$ | $1.3837 \times 10^{-22}$ | $5.0645 \times 10^{-23}$ | $5.9922 \times 10^{-23}$ | $4.9976 \times 10^{-24}$ | $2.2672 \times 10^{-23}$ | $2.1996 \times 10^{-22}$ |
| $D_{\rm H\&He}(T, n_{\rm H})$ | 0.96445 | 0.96839 | 0.99353 | 0.99999 | 1.0000 | 1.0000 |
| $M_{\rm H\&He}(T, n_{\rm H}, Z)$ | 1.0136 | 1.0637 | 1.0000 | 1.0000 | 1.0000 | 1.0000 |
| $\Lambda_{\rm metal}(T)$ | $2.1929 \times 10^{-23}$ | $5.1009 \times 10^{-23}$ | $2.5658 \times 10^{-22}$ | $9.4800 \times 10^{-23}$ | $4.5341 \times 10^{-24}$ | $1.6073 \times 10^{-23}$ |
| $D_{\rm metal}(T, n_{\rm H})$ | 0.70495 | 0.80362 | 0.98999 | 0.99999 | 1.0000 | 1.0000 |
| $\Lambda_{\rm ee}(T)$ | $2.3369 \times 10^{-31}$ | $5.1108 \times 10^{-31}$ | $3.6519 \times 10^{-30}$ | $1.8646 \times 10^{-28}$ | $4.8608 \times 10^{-25}$ | $1.2672 \times 10^{-21}$ |
| $M_{\rm ee}(T, n_{\rm H}, Z)$ | 1.0114 | 1.0157 | 1.0324 | 1.0790 | 1.1068 | 1.1078 |

Note. — This table lists the values of each fitting functions with $n_{\rm H} = 10^5 {\rm cm}^{-3}$ and $Z = 15 Z_\odot$. This table is only used to check whether user has used the fitting functions with correct fitting coefficients.